\documentclass[aps,pre,a4paper,twocolumn,amsmath,showpacs,floatfix]{revtex4}
\usepackage{graphicx}

\begin{document}
\bibliographystyle{apsrev}

\title{Phase diagrams of Zwanzig models: The effect of polydispersity}  

\author{Yuri Mart\'{\i}nez-Rat\'on}
\email{yuri@math.uc3m.es}
\author{Jos\'e A.~Cuesta}
\email{cuesta@math.uc3m.es}
\affiliation{Grupo Interdisciplinar de Sistemas Complejos (GISC), \\ 
Departamento de Matem\'aticas, Universidad Carlos III de Madrid, \\
Avda.~de la Universidad 30, E--28911, Legan\'es, Madrid, Spain.} 

\date{\today}

\begin{abstract}
The first goal of this article is to study the validity of the 
Zwanzig model for liquid crystals to predict transitions 
to inhomogeneous phases (like smectic and columnar) and the way
polydispersity affects these transitions. The second goal is to analyze
the extension of the Zwanzig model to a binary mixture of rods and
plates. The mixture is symmetric in that all particles have equal
volume and length-to-breadth ratio, $\kappa$. The phase diagram
containing the homogeneous phases as well as the spinodals of the
transitions to inhomogeneous phases is determined for the cases
$\kappa=5$ and 15 in order to compare with previous results obtained
in the Onsager approximation. We then study the effect of polydispersity
on these phase diagrams, emphasizing the enhancement of the stability
of the biaxial nematic phase it induces.
\end{abstract}

\pacs{64.70.Md,64.75.+g,61.20.Gy}

\maketitle

\section{Introduction}
\label{Introduccion}

Although the Zwanzig model was introduced long ago as a very simplified
model to study the isotropic-nematic transition in liquid crystals 
\cite{Zwanzig}, the determination of its phase diagram, including all 
inhomogeneous bulk phases, has never been carried out so far. With the
help of the scaled-particle theory (equivalent to a $y3$ expansion),
developed for a variety of hard particle fluids \cite{Barboy}
including the hard parallelepipeds with restricted orientations which
define the Zwanzig model \cite{Moore}, the phase behavior of the
homogeneous bulk phases has been investigated. Also the the
isotropic-nematic interface has been accessible to calculations
through a smoothed-density approximation \cite{Moore}, which
consists in evaluating the bulk free energy at some smoothed density,
following the same recipe of Tarazona for hard spheres \cite{Tarazona}.

The usual second virial approximation (exact for freely rotating models
but only approximate for restricted orientation models) has recently
been applied to the Zwanzig model in order to elucidate the bulk phase
behavior of a polydisperse mixture of hard rods \cite{Clarke}. With an
inhomogeneous version of this approximation different interfaces of a 
monodisperse hard rod fluid \cite{Roij}, as well as its polydisperse 
counterpart \cite{Martinez-Raton}, have been studied. 

Unlike for hard objects with free orientations---like spherocylinders,
which have been extensively simulated \cite{Bolhuis1}, there is only
one simulation of the Zwanzig model in a square lattice \cite{Casey},
for parallelepipeds with dimensions 
$5\times 1\times D$ (with $D=$1, rods, or 5, plates) \cite{Casey}. 
The main reason for this lack of simulation data is that any Monte Carlo 
movement which includes reorientation of parallelepipeds to any of the 
three directions has a very low acceptance ratio due to the huge overlap 
between particles as the length-to-breadth ratio and the density increase.

The second virial approximation of the free energy of the Zwanzig model 
has also been employed to investigate the phase diagram of symmetric
mixtures of rods and plates \cite{Van-Roij1}. Stimulated by theoretical 
calculations made in the early 70's \cite{Alben} which show that a 
binary mixture of rods and plates can stabilize a biaxial nematic phase 
(in which the symmetry axes of particles of different types point along
mutually perpendicular directions), van Roij and Mulder studied the
relative stability of this phase against a nematic-nematic phase
separation \cite{Van-Roij1}. The authors calculated the phase diagram 
for different length-to-breadth ratios $\kappa\equiv L/D$.  
They showed that for $\kappa>8.8$ the biaxial phase is stable in a
relative small window above a multicritical point. How the topology of
these phase diagrams is modified when a scaled-particle approximation
(which includes higher order virial coefficients) is used instead of
the second virial is one of the open questions which we will try to answer 
here. Another one is the influence of polydispersity in the phase behavior
of the mono and bidisperse Zwanzig model. This is most relevant since
with a recently derived a fundamental measure functional 
(FMF) for hard parallelepipeds \cite{Cuesta} (equivalent to the 
scaled-particle theory for homogeneous phases),
we have shown that polydispersity may enhance the
thermodynamic stability of a biaxial nematic phase
\cite{Martinez-Raton1}. 

%
There are 
recent experiments on true mixtures of hard rods and plates which show 
a very rich phase behavior, including triple coexistence between an 
isotropic and two nematics---each rich in one of the species---as well
as inhomogeneous phases, like a columnar phase \cite{Kooij}. This phase
behavior has been quantitatively accounted for using the Parson
approximation of the free energy functional of binary mixture of rods
and plates in the Onsager limit \cite{Wensink}. Nevertheless, the long 
searched biaxial phase has not been found in these experiments. 
An unavoidable ingredient of experiments is polydispersity. In
Ref.~\cite{Martinez-Raton1} we have shown that this otherwise undesirable
element may cause the stabilization of the biaxial nematic phase. It is
very important, though, that the system maintains rod-plate symmetry,
in contrast with what happens in the existing experiments 
\cite{Kooij}. An open question in this work is the topology of the phase 
diagram at high packing fractions. Addressing this problem requires to
study the possible phase transitions to inhomogeneous phases---like
smectic, columnar and solid phases. The calculation of phase equilibria
in polydisperse liquid crystals when one of the bulk phase is 
inhomogeneous is thus far a theoretical challenge. There is only one such
work, which makes use of an approximate functional for length-polydisperse
parallel hard cylinders to make two important predictions: the
existence of a terminal polydispersity beyond which the smectic phase 
is no more stable, and the enhancement of the columnar phase stability 
\cite{Bohle}. These results have been confirmed in simulations of 
freely rotating hard spherocylinders in the Onsager limit \cite{Bates}.

The paper is organized as follows. In Sec.~\ref{Model} we describe both
the model and the fundamental measure functional we will use to describe
its free energy (A); the equations for phase equilibria between phases
of a polydisperse system (B); and the formalism to determine spinodal
instabilities in mono, bi or polydisperse systems (C). Section \ref{results}
describes the phase diagrams of the Zwanzig model without (A) and with
polydispersity (B), as well as the bidisperse rod-plate model without
(C) and with polydispersity (D). Finally we conclude in
Sec.~\ref{conclusions}.

\section{Theory}
\subsection{Model}
\label{Model}

Let us consider a length-to-breadth polydisperse mixture of uniaxial 
oblate and prolate parallelepipeds, with their symmetry axes pointing
along one of the three coordinate axes. Let us fix the volume of any 
particle to 1; thus if  $\lambda=L/D$ is the length-to-breadth ratio
(with $L$ the length and $D$ the breadth) of a parallelepiped, then
\begin{equation}
L=\lambda^{2/3}, \quad D=\lambda^{-1/3}.
\label{scaling}
\end{equation}
Let us define $\rho^{\nu}(\lambda)$ to be the density distribution
function of the species parallel to the $\nu$ ($=x,y,z$) axis, and let
the total number density to be
\begin{equation}
\rho=\int_0^{\infty} d\lambda\,\rho(\lambda), \qquad
\rho(\lambda)=\sum_{\nu}\rho^{\nu}(\lambda).
\label{density}
\end{equation}

The (temperature reduced) free energy density of a multicomponent mixture
is given by $\beta FV^{-1}\equiv\Phi=\Phi_{\rm{id}}+\Phi_{\rm{exc}}$,
where $\Phi_{\rm{id}}$ is the ideal part, whose exact form is
\begin{equation}
\Phi_{\rm{id}}=\sum_{\nu,i}\rho_i^{\nu}\left(\ln
\rho_i^{\nu}-1\right),
\end{equation}
$i$ labeling different components, and the excess part, $\Phi_{\rm{exc}}$,
can be approximated by the FMF for hard parallelepipeds \cite{Cuesta},
namely
\begin{equation}
\Phi_{\rm{exc}}=-n_0\ln(1-n_3)+\frac{\sum_{\nu}n_1^{\nu}n_2^{\nu}}
{1-n_3}+\frac{\prod_{\nu}n_2^{\nu}}{(1-n_3)^2},
\end{equation}
where the weighted densities $\{n_{\alpha}\}$ have the form
\begin{eqnarray}
n_0=\sum_{\nu,i}\rho_i^{\nu}\equiv \rho, \quad n_1^{\nu}
=\sum_i \left[\rho D_i+\rho_i^{\nu}\left(L_i-D_i\right)\right], \\
n_3=\sum_{\nu,i}\rho_i^{\nu} L_i D_i^2, \quad 
n_2^{\nu}=\sum_i \left[\rho L_i -\rho_i^{\nu}\left(L_i-D_i\right)
\right]D_i. 
\end{eqnarray}
Specializing the above expressions for our polydisperse mixture and
using Eq.~(\ref{scaling}) we obtain
\begin{eqnarray}
\Phi_{\rm{id}}&=&\sum_{\nu}\int_0^{\infty} d\lambda \rho^{\nu}(\lambda)
\left[\ln \rho^{\nu}(\lambda)-1\right], \\
\Phi_{\rm{exc}}&=&-\rho \ln(1-\rho)+\frac{\sum_{\nu}\xi^{\nu}_-
\xi^{\nu}_+}{1-\rho}+\frac{\prod_{\nu}\xi_+^{\nu}}{(1-\rho)^2}, \\
\xi_{\pm}^{\nu}&=&\int_0^{\infty}\lambda^{\pm 1/3}\left[
\left(\lambda^{\mp 1}-1\right)\rho^{\nu}(\lambda)+\rho(\lambda)\right].
\label{xis}
\end{eqnarray}
The pressure can be obtained from its definition
\begin{eqnarray}
\beta\Pi &=& \sum_{\nu}\int_0^{\infty}d\lambda\,\left[\rho^{\nu}(\lambda)
\Phi^{\nu}(\lambda)\right]-\Phi, \\
\Phi^{\nu}(\lambda) &\equiv& \frac{\delta\Phi}{\delta\rho^{\nu}(\lambda)},
\end{eqnarray}
which in this particular case yields
\begin{equation}
\beta \Pi=\frac{\rho}{1-\rho}+\frac{\sum_{\nu}\xi_-^{\nu}\xi_+^{\nu}}
{(1-\rho)^2}+\frac{2\prod_{\nu}\xi_+^{\nu}}{(1-\rho)^3}.
\end{equation}

\subsection{Phase equilibria between homogeneous phases}
\label{phase}

To obtain the phase equilibria we follow the general procedure already 
reported elsewhere \cite{Clarke,Sollich}. Suppose that among the $n$ 
coexisting phases there are $n_I$ isotropic, $n_N$ nematic and $n_B$ 
biaxial phases ($n=n_I+n_N+n_B$). The global density distribution (the
parent distribution) is fixed to be 
\begin{equation}
P(\lambda)=P_0 h(\lambda), \quad \int d\lambda h(\lambda)=1,
\end{equation}
so total mass conservation of each species is expressed by the lever rule 
\begin{equation}
P(\lambda)=\sum_{a=1}^n\gamma_{a}\rho_{a}(\lambda), 
\label{level}
\end{equation}
with $\rho_{a}(\lambda)$ the total density distribution of phase 
$a$ and $\gamma_{a}$ the fraction of the total volume it 
occupies (there is the obvious constraint $\sum_{a}\gamma_{a}=1$).
Minimizing the free energy density $\Phi$ with respect to the fraction 
of particles with length-to-breadth ratio $\lambda$ oriented along the
$\nu$ axis, i.e.\ $p_{a}^{\nu}(\lambda)\equiv
\rho_{a}^{\nu}(\lambda)/\rho_{a}(\lambda)$, and using 
Eq.~(\ref{level}), the equality of the chemical potentials of each species
in different phases 
\begin{equation}
\beta \mu_{a}(\lambda)=
\sum_{\nu} \rho_{a}^{\nu}(\lambda)\Phi^{\nu}_{a}(\lambda),
\end{equation}
leads to the following expressions for the coexisting densities
\begin{equation}
\rho_{a}^{\nu}(\lambda)=P_0 h(\lambda)\frac{
e^{-\Phi^{\nu}_{a}(\lambda)}}{\sum_{b}\gamma_{b}
\sum_{\tau} e^{-\Phi^{\tau}_{b}(\lambda)}}
\label{distri}
\end{equation}

The parent distribution we are going to use throughout the paper is
characterized by
\begin{eqnarray}
h(\lambda)=\lambda^{-1}\left[\zeta f(\lambda/\kappa)+
(1-\zeta)f(\lambda\kappa)\right], \label{parent}\\
f(z)=K_0(\alpha)^{-1}\exp{[-(\alpha/2)\left(z^2+z^{-2}\right)]},
\end{eqnarray}
where $K_{\nu}(\alpha)$ $(\alpha>0)$ is the $\nu$th-order modified 
Bessel function, and $\kappa>1$. This choice is motivated by its 
rod-plate symmetry for $\zeta=1/2$. The best way to appreciate this 
symmetry is changing the variable to $\ln \lambda$: 
$\tilde{h}(\ln\lambda)=\lambda h(\lambda)$ has two identical humps 
(wider the smaller $\alpha$) centered at $\ln \kappa$ (rods of 
``typical'' aspect ratio $\kappa$) and $-\ln \kappa$ (plates of 
typical aspect ratio $\kappa^{-1}$). The parameter $0\le\zeta\le 1$
allows one to tune the overall composition of the mixture, since the molar 
fraction of the rods is given by $x_r(\zeta)=\int_1^{\infty} h(\lambda)
d\lambda$ [and that of plates by $x_p(\zeta)=1-x_r(\zeta)$, 
of course]. Thus we can select polydisperse rods (plates) by setting
$\zeta=1$ ($\zeta=0$). The moments of this distribution are given by 
$\langle \lambda^m\rangle=K_{m/2}(\alpha)K_0(\alpha)^{-1}
[\zeta \kappa^m+(1-\zeta)\kappa^{-m}]$, explicitly showing the 
symmetry of the mixture. A quantitative characterization of the 
polydispersity can be given if we determine the dispersion in $L$ and 
$D$ as obtained from $h(\lambda)$ for $\zeta=0$ or 1. This yields
\begin{equation}
\Delta_{L,D}\equiv\sqrt{\frac{\langle\lambda^{2\nu/3}
\rangle}{\langle\lambda^{\nu/3}\rangle^2}-1}=
\sqrt{\frac{K_{\nu/3}(\alpha)K_0(\alpha)}{K_{\nu/6}
(\alpha)^2}-1},
\end{equation}
where $\nu=2$ for $\Delta_L$ and $\nu=1$ for $\Delta_D$.

The number of independent moments in the set $\{\rho,\xi_{\pm}^{\nu}\}$  
which completely determine each coexisting phase is 3, 5 and 7 for the
isotropic, nematic and biaxial phases respectively (which amounts to a 
total of $3n_I+5n_N+7n_B$ unknowns). They can be obtained through
the definitions (\ref{density}) and (\ref{xis}), together with the 
distribution functions (\ref{distri}). The remaining unknowns---the 
$n-1$ independent $\gamma_{\alpha}$'s---are calculated from
the equality of pressures in every phase. This leaves the global dilution,
$P_0$, as the control parameter. Alternatively, we can fix an external
pressure, $\Pi_0$, and eliminate $P_0$ in terms of this new control
parameter. As $P_0$ increases, the fractions of volume of each phase
change; thus, for practical purposes, it is computationally simpler to
use one of the $\gamma_{\alpha}$'s as control parameter and obtain $P_0$
as a function of it.

A particularly important case is the two phase coexistence between a phase 
(say $\beta$) which fills the whole volume (cloud phase) and an incipient
new phase (say $\sigma$) which fills an infinitesimally amount of volume
(shadow phase). In a cloud-shadow coexistence the parent $P(\lambda)$ 
coincides with the distribution function of the cloud phase. Then $\rho$
and $\sum_{\nu}\xi^{\nu}_{\pm}$ are fixed for this phase, so the number 
of unknowns reduces by 3 (the number of constraints).

\subsection{Spinodal instabilities with respect to inhomogeneous phases}

In order to be as general as possible in developing the formalism
let us consider the problem of finding the conditions for the
stability of an arbitrary multicomponent system against spatial
modulations of the densities $\rho_i$ ($i$ label the species).
This way we can replace at the end $\sum_i\rightarrow\sum_{\nu}$ 
for the monodisperse Zwanzig model and $\sum_i\rightarrow\sum_{\nu}
\int d\lambda$ for the polydisperse one, to obtain the respective
spinodal instabilities. The reason for so proceeding is that, as we
will show later, the spinodal equations can be expressed in two
different ways, each of which suits one of the two cases.

The condition of equality of chemical potentials between two 
phases is equivalent to the minimization of the grand potential
\begin{equation}
\Omega=\int d{\bf r}\,\left\{k_{\rm B}T\Phi(\{n_{\alpha}\})-
\sum_i\mu_i\rho_i({\bf r})\right\},
\label{grand}
\end{equation}
at fixed chemical potentials of each species $\mu_i$, with
respect to the density profile of each species $\rho_i({\bf r})$.
We specify this grand potential for a general model whose free energy 
density $\Phi$ has an excess part depending on the densities only 
through certain weighted densities 
\begin{equation}
n_{\alpha}({\bf r})=\sum_i \left[\rho_i\ast \omega^i_{\alpha}\right]
({\bf r}),
\end{equation}
as it is the case of Rosenfeld's fundamental measure theory 
\cite{Rosenfeld}. In the 
definition of $n_{\alpha}$ the symbol ``$\ast$'' stands for the 
convolution of two functions, i.e.\ $f\ast g({\bf r})\equiv
\int d{\bf r}'\,f({\bf r}')g({\bf r}-{\bf r}')$. 

We are concerned with the case when one of the coexisting phases is 
homogeneous and the other one is inhomogeneous, with its density profile 
consisting in a small perturbation around the homogeneous phase with 
density distribution $\rho_i$, i.e.
\begin{equation}
\rho_i({\bf r})=\rho_i\left[1+\epsilon_i({\bf r})\right], 
\qquad \epsilon_i({\bf r})\ll 1.
\end{equation}
The result of minimizing (\ref{grand}) with respect to 
$\rho_i({\bf r})$ can be cast as
\begin{equation}
\rho_i({\bf r})=\rho_i\exp\left\{-\sum_{\alpha} 
\left[\Delta\phi_{\alpha} \ast \omega^i_{\alpha}\right]({\bf r})
\right\},
\label{equi}
\end{equation}
where $\Delta \phi_{\alpha}({\bf r})=
\phi_{\alpha}({\bf r})-\phi_{\alpha}$ and
$\phi_{\alpha}=\partial \phi/\partial n_{\alpha}$. We are implicitly
using magnitudes without spatial variable arguments to denote homogeneous 
phase quantities. Expanding $\Delta\phi_{\alpha}({\bf r})$ to first
order in $\epsilon_i({\bf r})$ yields
\begin{equation}
\Delta \phi_{\alpha}({\bf r})=
\sum_{\beta}\phi_{\alpha \beta}\sum_j\rho_j
\left[\epsilon_j\ast \omega^j_{\beta}\right]({\bf r})+\cdots,
\label{series}
\end{equation}
where $\phi_{\alpha\beta}=\partial^2 \Phi/(\partial n_{\alpha}\partial
n_{\beta})$.
Inserting (\ref{series}) in (\ref{equi}) and expanding the exponential 
again to first order in $\epsilon_i({\bf r})$ we arrive at the 
integral equation
\begin{equation}
\epsilon_i({\bf r})=-\sum_{\alpha,\beta}\phi_{\alpha\beta} 
\sum_j \rho_j \left[\epsilon_j \ast
\omega^i_{\alpha}\ast\omega^j_{\beta}\right]({\bf r}),
\label{pert}
\end{equation}
which is handier to write in Fourier space
\begin{equation}
\varepsilon_i({\bf q})+\sum_{\alpha,\beta}\phi_{\alpha\beta}
\Omega^i_{\alpha}({\bf q})
\sum_j \rho_j\varepsilon_j({\bf q})
\Omega^j_{\beta}({\bf q})=0,
\label{fourier}
\end{equation}
where $\varepsilon$ and $\Omega$ are respectively the Fourier 
transforms of $\epsilon$ and $\omega$ [with the usual definition 
$f({\bf q})=\int d{\bf r} e^{i{\bf q}{\bf r}}f({\bf r})$]. 

Equation (\ref{fourier}) gets clearer if written in matrix form.
Defining the functions
\begin{equation}
m_{ij}({\bf q})\equiv \sum_{\alpha,\beta}\phi_{\alpha\beta}
\Omega^i_{\alpha}({\bf q})\Omega^j_{\beta}({\bf q}),
\end{equation}
the matrices $\hat{\rm M}({\bf q})\equiv\big(m_{ij}({\bf q})\big)$
and $\hat{\rm P}\equiv\big(\rho_j\delta_{ij}\big)$, and the vector
$\boldsymbol\varepsilon({\bf q})\equiv\big(\varepsilon_i({\bf q})\big)$,
Eq.~(\ref{fourier}) becomes
\begin{equation}
\left({\rm I}+\hat{\rm M}({\bf q})\cdot\hat{\rm P}\right)
\boldsymbol\varepsilon({\bf q})=0,
\end{equation}
where I is the identity matrix.
This linear system has nontrivial solutions if and only if
\begin{equation}
D(\rho,{\bf q})\equiv{\rm det}\left[{\rm I}+
\hat{\rm M}({\bf q})\cdot\hat{\rm P}\right]=0.
\label{det1}
\end{equation}
This equation is suitable for the cases in which there is a small
number of components, like the monodisperse Zwanzig model with
only three components: $\nu=x,y,z$.

Alternatively we can multiply (\ref{fourier}) by 
$\rho_i\Omega^i_{\gamma}({\bf q})$, sum over $i$ and define
the functions
\begin{equation}
u_{\gamma}({\bf q})\equiv\sum_i \rho_i
\varepsilon_i({\bf q})\Omega^i_{\gamma}({\bf q});
\end{equation}
then Eq.~(\ref{fourier}) becomes
\begin{equation}
u_{\gamma}({\bf q})+\sum_{\alpha,\beta}\phi_{\alpha\beta}
n_{\alpha\gamma}({\bf q})u_{\beta}({\bf q})=0,
\label{final}
\end{equation}
where we have introduced new functions
\begin{equation}
n_{\alpha\gamma}({\bf q})=\sum_i \rho_i
\Omega^i_{\alpha}({\bf q})\Omega^i_{\gamma}({\bf q}).
\end{equation}
Equation (\ref{final}) can also be written in matrix form, but
now the indices run over the set of weights, not
the components. Hence, defining the matrices $\hat{\Phi}\equiv
\big(\phi_{\alpha\beta}\big)$ and $\hat{\rm N}({\bf q})\equiv
\big(n_{\alpha\gamma}({\bf q})\big)$, and the vector
${\bf u}({\bf q})\equiv\big(u_{\alpha}({\bf q})\big)$, Eq.~(\ref{final})
becomes
\begin{equation}
\left[{\rm I}+\hat{\rm N}\cdot \hat{\Phi}\right]{\bf u}={\bf 0}.
\label{matrix}
\end{equation}
Again this system has nontrivial solutions if and only if 
\begin{equation}
D(\rho,{\bf q})\equiv{\rm det}\left[{\rm I}+
\hat{\rm N}\cdot \hat{\Phi}\right]=0.
\label{det}
\end{equation}
This alternative characterization of the spinodal is suitable
for polydisperse systems (actually, it is a generalization of the
formalism developed in Ref.~\cite{CuestaEPL} for homogeneous
phases). The reason is that it replaces integral operators by finite
matrices, in which the number of components is limited by the number
of weights of the theory. For instance, for the polydisperse Zwanzig
model with a free energy functional given by the FMF derived in 
Ref.~\cite{Cuesta}, functions $n_{\alpha\beta}$ take the form
\begin{equation}
n_{\alpha\beta}({\bf q})=\sum_{\nu}\int d\lambda\,\rho^{\nu}(\lambda)
\Omega_{\alpha}^{\nu}({\bf q},\lambda)\Omega_{\beta}^{\nu}
({\bf q},\lambda),
\end{equation}
where $\alpha,\beta=0,(1x,1y,1z),(2x,2y,2z),3$ and
\begin{eqnarray}
\Omega_0^{\nu}({\bf q},\lambda) &=& \prod_k w_0(q_k\Lambda_{\nu k}), 
\label{omega0}\\
\Omega_3^{\nu}({\bf q},\lambda) &=& \prod_k \Lambda_{\nu k}
w_3(q_k\Lambda_{\nu k}), \\
\Omega_{1,k}^{\nu}({\bf q},\lambda) &=&
\Lambda_{\nu k}w_1(q_k\Lambda_{\nu k})
\Omega_0^{\nu}({\bf q},\lambda), \label{omega1}\\ 
\Omega_{2,k}^{\nu}({\bf q},\lambda) &=&
[\Lambda_{\nu k}w_1(q_k\Lambda_{\nu k})]^{-1}
\Omega_3^{\nu}({\bf q},\lambda) \label{omega2}
\label{omega3}
\end{eqnarray}
($k=x,y,z$). Here $q_k$ are the components of the vector ${\bf q}$,
and
\begin{eqnarray}
\Lambda_{\nu k} &=& \lambda^{-1/3}+(\lambda^{2/3}-\lambda^{-1/3})
\delta_{\nu k}, \\
w_0(x) &=& \cos(x/2), \\
w_3(x) &=& 2\sin (x/2)/x, \\
w_1(x) &=& w_3(x)/w_0(x).
\end{eqnarray}
As there are  
8 different weights, $D(\rho,{\bf q})$ is in this case the determinant
of an $8\times 8$ matrix. 

Either way we express the equation, (\ref{det1}) or (\ref{det}),
the spinodal instability condition
of a homogeneous phase with density distributions $\rho^{\nu}(\lambda)$ 
(or simply $\rho^{\nu}$ in the monodisperse case)
with respect to the transition to an inhomogeneous phase occurs at
the values of the total density $\rho$ and wave vector ${\bf q}$ for 
which the function $D(\rho,{\bf q})$ first vanishes. More precisely, 
this function has 
an oscillatory behavior as a function of  ${\bf q}$, and is positive 
as long as the homogeneous phase is stable, so the spinodal corresponds
to the smallest density $\rho$ for which the absolute minimum of $D$ 
with respect to ${\bf q}$ equals zero. This amounts to finding, for
a given $\rho$, the solutions to
\begin{equation}
\boldsymbol{\nabla} D(\rho,{\bf q})=0.
\label{nabla}
\end{equation}

When we later consider the rod-plate mixture we will have to locate
the nematic-biaxial phase transition. As it is a continuous transition
it can also be found as a solution to Eq.~(\ref{det}) with ${\bf q}=0$
(both phases are homogeneous), using
\begin{equation}
\rho^{\nu}(\lambda)=P_0h(\lambda)\frac{e^{-\Phi^{\nu}(\lambda)}}
{\sum_{\tau} e^{-\Phi^{\tau}(\lambda)}}
\label{cloud}
\end{equation}
as the distribution functions of the nematic phases. 

\section{Results}
\label{results}
\subsection{Monodisperse Zwanzig model}
\label{one}

As a first step towards the calculation of the whole phase diagram  
of the monodisperse (pure rods or pure plates) Zwanzig model
using the FMF of Ref.~\cite{Cuesta} we will determine the 
isotropic-nematic (I--N$^{\pm}$) coexistence curves as well
as the spinodals for the nematic-smectic (N$^{\pm}$--S$^{\pm}$), 
nematic-columnar (N$^{\pm}$--C$^{\pm}$) and isotropic-plastic 
(orientationally disordered) solid (I--PS)
transitions, as a function of the length-to-breadth ratio $\kappa$.
The $+$ and $-$ superscripts label prolate (rods) and oblate (plates)
parallelepipeds respectively. If the transitions are first order,
the location of the coexistence curves will differ
from that of the corresponding spinodals. However, as
the main concern of this paper is the effect of polydispersity, we
defer the calculations of coexistence with and between inhomogeneous
phases to a forthcoming publication.

We have obtained the spinodals for this model by solving 
Eqs.~(\ref{det1}) and (\ref{nabla}) using the monodisperse version
of (\ref{cloud}). If we choose the nematic director parallel to the
$z$ axis, then the N$^{\pm}$--S$^{\pm}$ spinodals can be calculated
setting ${\bf q}=(0,0,q)$, whereas the N$^{\pm}$--C$^{\pm}$ ones
follow from taking ${\bf q}=(q,0,0)$ or ${\bf q}=(0,q,0)$ 
(both are equivalent due to the nematic symmetry). For the 
I--PS spinodal all three previous vectors give the same result. 

\begin{figure}
\mbox{\includegraphics*[width=3.in, angle=0]{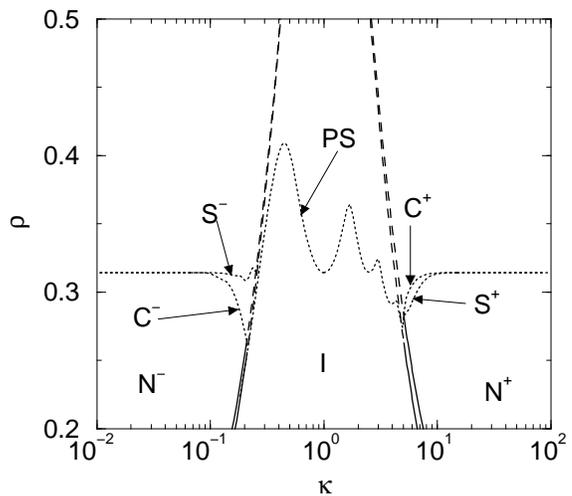}}
\caption[]{Phase diagram (packing fraction, $\rho$, vs.\
length-to-breadth ratio, $\kappa$) of the monodisperse Zwanzig model. 
Solid and dashed lines represent the isotropic (I)-nematic (N$^{\pm}$)
coexistence curves. The dashed lines correspond to the values for which
the homogeneous 
phases are unstable with respect to the inhomogeneous ones. Dotted lines
represent the spinodals of the different homogeneous phases: each one is
labeled with the corresponding phase: smectic (S$^{\pm}$), columnar 
(C$^{\pm}$) or plastic solid (PS).}
\label{fig1}
\end{figure}

Figure~\ref{fig1} shows the results of these calculations. We observe in
this figure that there is I--N$^{\pm}$ transition only for $\kappa<0.210$
and $\kappa>5.02$. At those two limiting values of $\kappa$ the N$^+$ phase
is destabilized by a S$^+$ at a packing fraction $\rho=0.280$, and the
N$^-$ by a C$^-$ at a packing fraction $\rho=0.261$ (remember the particle
volume has been set to 1). Also, the N$^+$--S$^+$ spinodal is always 
below the N$^+$--C$^+$, and the N$^-$--C$^-$ spinodal is always below
the N$^-$--S$^-$ one. This is what intuition tells us but, as we will see
later, it is a peculiarity of the monodisperse system. Despite this,
all these spinodal curves converge asymptotically to 
the same packing fraction, $\rho=0.314$, as $\kappa\to\infty$ or 
$\kappa\to 0$. This is precisely the packing fraction at which the 
continuous freezing transition occurs in a system of parallel hard cubes
($\kappa=1$) \cite{yuri1}. The reason for this is that
upon increasing $\kappa$ the number of rods with orientation perpendicular
to the director becomes vanishing small, and then the system is, after
rescaling the $z$ direction, almost equivalent to a system of parallel
cubes. The same holds for plates upon decreasing $\kappa$. What this means
is that most likely the N$^+$--S$^+$ and N$^-$--C$^-$ transitions will
be metastable with respect to freezing in a large portion of the phase
diagram. For $0.182<\kappa<4.93$ we have found an
I--PS spinodal instability. The
peculiarity of this curve is that it exhibits strong oscillations as the
aspect ratio changes. These oscillations reflect the packing efficiency
of randomly oriented parallelepipeds as a function of their size (the 
better the packing the lower the curve). 

The available simulation results for freely rotating hard spherocylinders
show that the I--S$^+$ and N$^+$--S$^+$ transitions begin at $\kappa=4.1$ 
and $4.5$ respectively \cite{Bolhuis1} (notice that for hard spherocylinders
the length-to-breadth ratio is $\kappa=\frac{L+D}{D}$).
On the other hand, simulations
of hard cut spheres show that for $\kappa=0.2$ there is an I--C$^-$
transition (the isotropic phase might instead be a peculiar `cubatic' phase)
and for $\kappa=0.1$ a N$^-$--C$^-$ one \cite{Veerman}. We can see that
despite the different particle geometry and the restricted orientations 
of the Zwanzig model,
the agreement with the threshold $\kappa$'s at which spatial instabilities
destabilize the homogeneous phases is rather good. Also the qualitative
picture is similar: elongated rods form smectics, while flat disks form
columnars, and more symmetric particles form solids instead of smectics or
columnar. Thus, this simple model seems to capture the essence of 
the entropy driven phase transitions between phases with different 
symmetries and their relation to particle anisotropy.

The only existing simulation of the Zwanzig model has been performed
on a lattice model of parallelepipeds with length-to-breadth ratios 
$15:3$ ($\kappa=5$) and $3:15$ ($\kappa=0.2$) \cite{Casey}. For these 
$\kappa$'s the authors find I--C$^-$ for the discs and and I--DS$^+$ 
for the rods, where DS$^+$ stands for a novel phase called `discotic
smectic' by the authors (in this phase the axes of the particles point 
perpendicular to the normal of the smectic layers while there is no
orientational order within the layer). The packing fractions they obtain
for the transitions are higher than those of the corresponding spinodals 
of Fig~\ref{fig1}, but they should decrease upon decreasing the lattice
spacing, as the results for the freezing of parallel
hard cubes on a lattice (occurring at $\rho=0.568$ for edge-length 2
lattice spacings, at $\rho=0.402$, for edge-length 6 lattice spacings,
and at $\rho=0.314$ for the continuum) illustrate \cite{Lafuente}. All
this is again in agreement with the phase diagram of Fig.~\ref{fig1}.

\subsection{Polydisperse Zwanzig model: Unimodal distribution}
\label{poly}

Let us study now the effect polydispersity has on the phase 
behavior of the Zwanzig model discussed above. 
We have introduced polydispersity through the unimodal 
parent probability density resulting when setting
$\zeta=1$ (rod-like parallelepipeds) or $\zeta=0$ (plate-like 
parallelepipeds) in function (\ref{parent}). Before reporting the results
we have obtained, a few words on polydisperse phase diagram plots
are on purpose.

Generally speaking, plotting polydisperse phase diagrams would require
an infinite dimensional space, for as we vary the fraction of total
volume occupied by the coexisting phases they change their composition.
A phase diagram like that of Fig.~\ref{fig1} plots densities vs.\ 
aspect ratios, thus carrying no information whatsoever about compositions.
So we must restrict ourselves to a given composition, and this is the
parent one, since all pure phases have this composition. This means
that only cloud lines can be plotted, delimiting regions where we can
only know that the system is decomposed into two or more coexisting phases.
The result resembles the coexisting lines of monodisperse phase diagram,
but the meaning is completely different. Although we could connect cloud
points in the two lines delimiting the coexistence region, they
are definitely no coexisting states.

Similar considerations hold for spinodal lines. We can represent the 
spinodal of a system in a pure phase having the parent distribution as
composition, but if the system reaches a cloud point an incipient shadow
phase with a totally different composition---hence a totally different
spinodal line---coexists with it. If the density of this shadow phase
is above its own spinodal line, the phase transition will not be stable.

\begin{figure}
\mbox{\includegraphics*[width=3.in, angle=0]{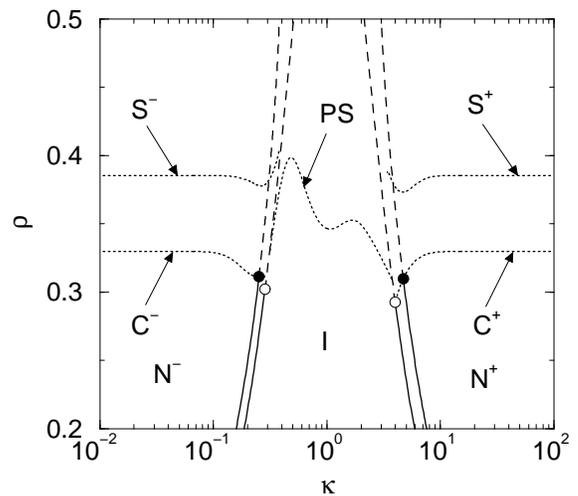}}
\caption[]{Phase diagram for polydisperse Zwanzig model. 
The distribution of aspect ratios is unimodal with the maximum at $\kappa$.
Polydispersity is $\Delta_{L}=0.288$ in length and $\Delta_{D}=0.143$ 
in breadth. The lines have the same meanings as in Fig.~\ref{fig1}.
Full circles mark the points above which the nematic cloud phase is
unstable against columnar ordering; empty circles mark the points above
which the isotropic cloud phase coexists with a nematic shadow phase
unstable against columnar modulations (the spinodal line of this nematic
shadow, computed with the corresponding aspect ratio distribution,
is not plotted in the figure).}
\label{fig2}
\end{figure}

Having all this in mind we have solved the coexistence equations
of the isotropic and nematic phases and determined the spinodal lines
of inhomogeneous instabilities for a system with the parent distribution
(\ref{parent}). The results for the choice $\alpha=1$ (corresponding
to polydispersity parameters $\Delta_{L}=0.288$ and $\Delta_{D}=0.143$)
are shown in Fig.~\ref{fig2}.

Apart from the usual broadening of the isotropic-nematic transition already 
observed for this model in Ref.~\onlinecite{Clarke}, we find a clear
enhancement of the stability of the homogeneous phases i.e.\ the  
spinodal instabilities occur at higher densities. This effect is more 
pronounced for the N$^{\pm}$--S$^{\pm}$ or N$^{\pm}$--C$^{\pm}$ transitions.
Another remarkable feature of Fig.~\ref{fig2} is that the columnar phases
are by far more stable than the smectic ones for any aspect ratio.

This general behavior is in agreement with the results obtained from 
both density functional theory for length-polydisperse parallel hard
cylinders \cite{Bohle} and computer simulations for length-polydisperse
freely rotating hard spherocylinders in the Onsager limit \cite{Bates}.
Although in Ref.~\cite{Bohle} some strong approximations were made, such as
the decoupling of the density profile as $\rho({\bf r},L)=g(L)\rho({\bf r})$,
and the effect of fractionation between the nematic and columnar phases 
was ignored, the results agree qualitatively with the simulations: both
show a terminal polydispersity beyond which the smectic phase is no more 
stable, being replaced by a columnar phase (which tolerates a higher degree 
of polydispersity). The nematic-columnar transition is first order. While 
fractionation is indeed negligible in the nematic-smectic transition,
the smectic-columnar transition clearly segregates long rods into the
columnar phase. 


It is interesting to notice that in the limits $\kappa\to \infty$ or 0 none
of the spinodal lines tend to the density of the freezing of
parallel hard cubes ($\kappa=1$). The reason is that no trivial scaling
can be applied to a polydisperse system of perfectly aligned parallelepipeds
to transform it into a system of parallel hard cubes. On the other hand
the S$^+$ and S$^-$ spinodals tend to the same limiting density when
$\kappa\to 0,\infty$, and so do the C$^+$ and C$^-$ spinodals. The second
virial term of the free energy is dominant for large (small) $\kappa$'s;
this and the rod-plate symmetry of the chosen parent distributions
explains this behavior.

\subsection{Binary mixture of Zwanzig rods and plates}
\label{binary}
\begin{figure}
\mbox{\includegraphics*[width=3.in, angle=0]{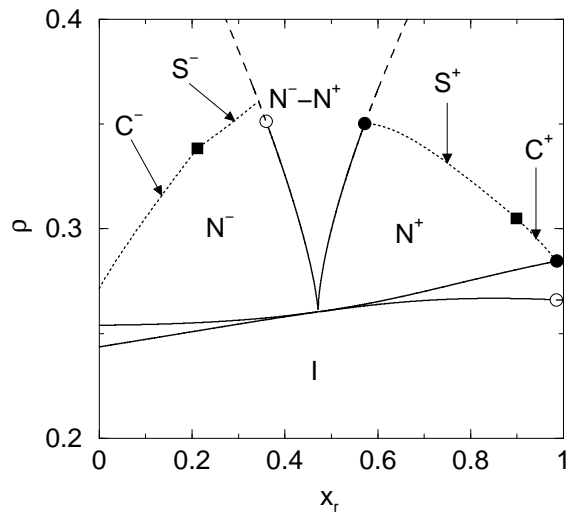}}
\caption[]{Phase diagram of a binary mixture of rods and plates with 
$\kappa=5$; $x_r$ is the fraction of rods and $\rho$ coincides with 
the packing fraction. Spinodal and coexistence curves are represented
by dotted and solid lines, respectively. Dashed lines are coexistence
curves in the unstable region. The crossings between the coexistence
curves and the spinodals are represented with full circles, and with
empty circles their coexisting phases. Full squares mark the crossover
between columnar and smectic phases along the spinodals.}
\label{fig3}
\end{figure}

We have calculated the phase diagram (coexistence of homogeneous 
phases as well as their spinodal instabilities to inhomogeneous phases) 
for the binary mixture of rods and plates with $\kappa=5$ and 15. The
results for $\kappa=5$ are shown in Fig.~\ref{fig3}. Apart from the usual
I--N$^{\pm}$ transitions and N$^-$--N$^+$ phase separations which coalesce
in a multicritical point we see that the nematic phases increase
their stability with respect to the inhomogeneous ones as the relative  
composition of the fluid tends to the equimolar mixture. Another important 
feature is that the biaxial nematic is not stable.

All these characteristics
are also present in the phase diagram obtained using a second virial 
approximation \cite{Van-Roij1}, but there are important differences as well.
The most prominent one is that the whole phase diagram is shifted down in
density, with all transitions and spinodals appearing below $\rho=0.4$
(those obtained with a second virial approximation occur above that
value \cite{Van-Roij1}). This result should not surprise if one takes
into account that in a $y3$ expansion the excess free energy is written
in terms of the variable $y=\rho/(1-\rho)$, which grows faster than
$\rho$. Another striking difference with respect to the second virial
approach is the rod-plate asymmetry of the phase diagram, a consequence
of the inclusion of higher virial coefficients---which do not share
the symmetry of the second one.

We pass now to describe the loss of stability of the nematics with 
respect to inhomogeneous phases. The intersection between the N$^+$ line 
of the I--N$^+$ coexistence and the N$^+$--C$^+$ spinodal in the rod-rich
part of the phase diagram (marked in the figure with a full circle;
the open circle corresponds to its coexisting isotropic phase) is
consistent with the existence of a first order phase transition between
the I phase and an inhomogeneous (presumably columnar) phase.
(Recall that simulations on the lattice find a first order I-DS$^+$
transition for parallelepipeds with $\kappa=5$ \cite{Casey}).
Nevertheless a definitive answer about the relative stability between
these inhomogeneous phases---including also the solid---can only be
given after carrying out coexistence calculations. 

Increasing the fraction of plates (which in the C$^+$ phase have their 
principal axes oriented perpendicular to the director) seems to favor 
the smectic alignment of rods. This results in the intersection of
the N$^+$--C$^+$ and the N$^+$--S$^+$ spinodals (marked with a full square 
on the right of Fig.~\ref{fig3}). In the second virial approach the
N$^+$--S$^+$ spinodal is always below the N$^+$--C$^+$ one. At the
plate-rich part of the phase diagram the nematic exhibits a spinodal
to the C$^-$ phase, as usual in discotic fluids \cite{Veerman}. 
Increasing the fraction of rods beyond a threshold value (indicated with
a full square on the left part of the phase diagram) makes the smectic
more stable than the columnar, which agrees
qualitatively with the results from the second virial 
approach \cite{Van-Roij1}.

\begin{figure}
\mbox{\includegraphics*[width=3.in, angle=0]{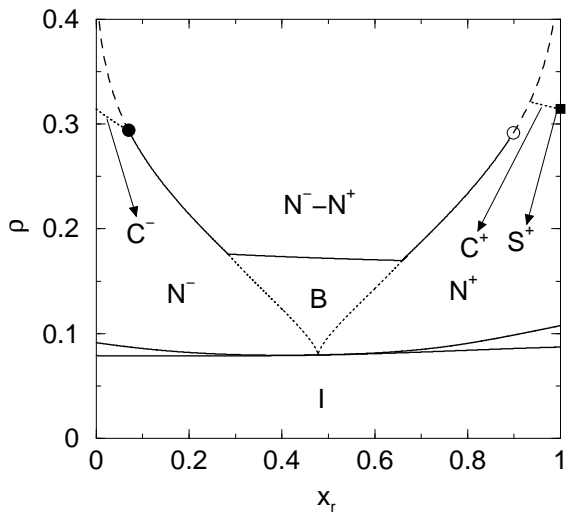}}
\caption[]{Phase diagram of binary mixture of rods and plates with 
$\kappa=15$. Lines and symbols have the same meanings as in
Fig.~\ref{fig3}. Notice that given the continuous nature of the
B--N$^{\pm}$ transitions, the spinodals are the true transition
lines.}
\label{fig4}
\end{figure}

The phase diagram for $\kappa=15$ appears in Fig.~\ref{fig4}. Its main
difference with the previous one is the presence of a
thermodynamically stable biaxial nematic phase in an inverted triangular
right above the multicritical point. The window is limited to the left
and to the right by continuous phase transitions to uniaxial nematic
phases, and to the top by a B--N$^+$ coexistence which
becomes a N$^-$--N$^+$ phase separation as density increases. As discussed
in Ref.~\cite{Van-Roij1}, the driving mechanism which determines the
preference of this system for phase separation instead of biaxial ordering
is the larger exclude volume between unlike particles compared to that
of like particles (the rod-plate excluded volume divided by the rod-rod
one scales as $\kappa^{2/3}$ for large $\kappa$). When the gain in free
volume compensates the loss in mixing entropy (and this strongly depends
on concentration and composition) phase separation occurs.

The slight asymmetry of the phase diagram is again due to the presence
of higher virial terms in the free energy. The second virial approximation
predicts a small region of triple coexistence between the N$^-$, N$^+$ and
B phases right on top of the biaxial phase region. The absence of this
triple coexistence region in Fig.~\ref{fig4} is due to the asymmetry of
the phase diagram. Simulations of a mixture of prolate and oblate 
ellipsoids \cite{Camp} show a similar asymmetric phase diagram, only tilted
in the opposite direction (there is a B--N$^-$ demixing on top of the biaxial
nematic, instead of the B--N$^+$ of the present model), probably due to
the different particle geometry. Despite the asymmetry, we should admit
that for $\kappa=15$ there is good overall agreement between the second
virial approach and ours.

We have not studied the changes an asymmetry in the mixture (a difference
in the volume of rods and plates, for instance) would induce in the
phase diagram. Intuitively, asymmetry acts against the stability of the
biaxial phase, something that is confirmed by a recent work
\cite{Wensink1} which studies a binary rod-plate mixture of freely
rotating oblate and prolate cylinders in the Onsager approximation. This
work explicitly shows that the region of stability of the biaxial 
phase decreases with increasing the asymmetry of the mixture (although,
according to this work, this phase can be found in highly asymmetry
mixtures).

With respect to the inhomogeneous phases, Fig.~\ref{fig4} shows a
remarkable behavior: although pure hard rods in the N$^+$ phase
undergo a spinodal instability to a smectic one, by adding a tiny
fraction of plates the instability takes place to a columnar phase.
This phenomenon has already been observed in a different system, namely,
in simulations of binary mixtures of parallel hard spherocylinders
\cite{Stroobants}. Although pure parallel hard spherocylinders exhibit
a continuous nematic-smectic transition, a binary mixture of them
with length-to-breadth ratios 2 and 2.9 show a first order nematic-columnar
transition instead. This result was explained by the poorer packing
of rods of different length in the smectic phase as compared to that
of rods of the same length.
Borrowing the argument, in our rod-plate model the plates
can fit into the interlayer spacing with their axes parallel
to the smectic director as long as there are few of them; but upon
increasing their molar fraction some of them are forced to get into 
the smectic layers with their axes perpendicular to the smectic director,
thereby destabilizing the smectic phase.

Finally, at the other extreme
of phase diagram the N$^-$ phase losses its stability to a columnar
phase of plates at densities lower than those of the
N$^+$--C$^+$ spinodal.

\subsection{Polydisperse Zwanzig model: Bimodal distribution}

In this section we are going to study the effect of polydispersity on
the phase behavior of the previous rod-plate mixture. In 
Ref.~\cite{Martinez-Raton1} we have shown how polydispersity can 
stabilize the biaxial phase even for mixtures with relative small 
aspect ratios---like $\kappa=5$, for which Fig.~\ref{fig3} shows
that the biaxial phase is absent.

Apart from the increase in mixing
entropy that polydispersity carries, the other mechanism behind the
enhancement of stability of the biaxial phase is the decrease of
the ratio of the average exclude volumes of like and unlike particles.
The excluded volume between two parallel rods
with lengths $L_i$ and breadth $D_i$ ($i=1,2$) is
\begin{equation}
v_{rr}=\left(L_1+L_2\right)\left(D_1+D_2\right)^2,
\end{equation}
and the excluded volume between a rod with length $L_1$ and breadth $D_1$
and a plate with length $L_2$ and breadth $D_2$, with axes
perpendicular to each other, is
\begin{eqnarray}
v_{rp}=\left(L_1+D_2\right)\left(L_2+D_1\right)
\left(D_1+D_2\right).
\end{eqnarray} 
According to (\ref{scaling}), if the volume of all particles is set to
one, in terms of the length-to-breadth ratios $\lambda_i$,
$L_i=\lambda_i^{2/3}$ and $D_i=\lambda_i^{-1/3}$. Taking a
double average of $v_{rr}$ over $\prod_i\left[h(\lambda_i)
\Theta(\lambda_i-1)\right]$ and of $v_{rp}$ over
$\left[\prod_i h(\lambda_i)\right]\Theta(\lambda_1-1)\Theta(1-\lambda_2)$
[$h(\lambda)$ is the parent distribution (\ref{parent})],
and fixing $\zeta=1/2$ (the equimolar composition) we arrive at
\begin{equation}
\frac{\langle\langle v_{rp}\rangle\rangle}{
\langle\langle v_{rr}\rangle\rangle}
=\frac{\frac{1}{2}+m_{-1}^2+m_1^2+2(m_{-1}m_2+m_{1}m_{-2})}{
\frac{1}{2}+4m_{-1}m_{1}+2m_{-2}m_{2}},
\label{ratio}
\end{equation}
where $m_{\beta}\equiv\int_{1}^{\infty}d\lambda h(\lambda)\lambda^{\beta/3}$.
It can be shown numerically that, for a given $\kappa$, the ratio 
(\ref{ratio}) decreases with polydispersity (i.e.\ with decreasing
$\alpha$). Analytic expressions for this ratio can be obtained in the
limit of high particle anisotropy ($\kappa\gg 1$) and high polydispersity
($\alpha\ll 1$) with the constraint $\alpha\kappa^2\ll 1$ (in terms of
the parameters $\Delta_\nu$, $\nu=L,D$, this constraint is equivalent
to $1\ll\Delta_{\nu}\ll\sqrt{\ln\kappa}$, implying that the fraction of
cubic-like particles is vanishing small). Using the asymptotic expressions
\begin{equation}
m_{\beta}\sim \frac{\kappa^{\beta/3}}{\alpha^{\beta/6}(-\ln\alpha)}, 
\qquad \Delta_{\nu} \sim \left(-\ln \alpha\right)^{1/2},
\end{equation}
we obtain
\begin{equation}
\frac{\langle\langle v_{rp}\rangle\rangle}{\langle\langle v_{rr}\rangle\rangle}
\sim \kappa^{2/3} e^{-c_{\nu} \Delta_\nu^2}, \quad \nu=L,D
\end{equation}
with $c_{\nu}$ a positive constant. This asymptotic relationship 
explicitly shows both the scaling of this ratio discussed for the
pure rod-plate mixture \cite{Van-Roij1} and the exponential attenuation
of this scaling with polydispersity.

\begin{figure}
\hspace*{6mm}\mbox{\includegraphics*[width=3.in, angle=0]{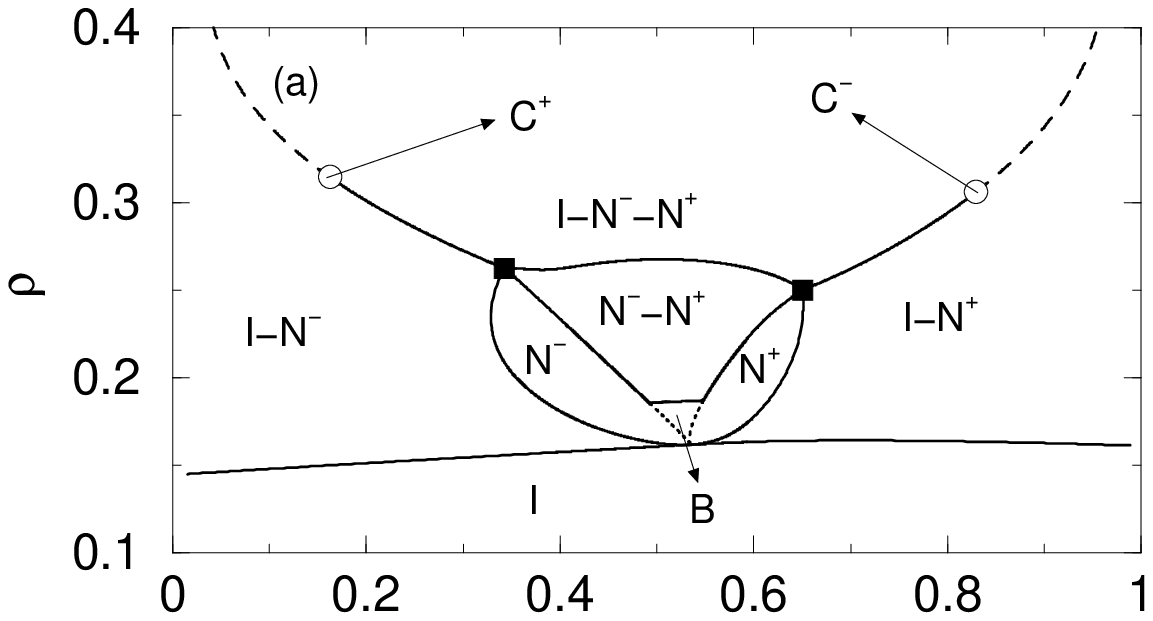}}
\mbox{\includegraphics*[width=3.4in, angle=0]{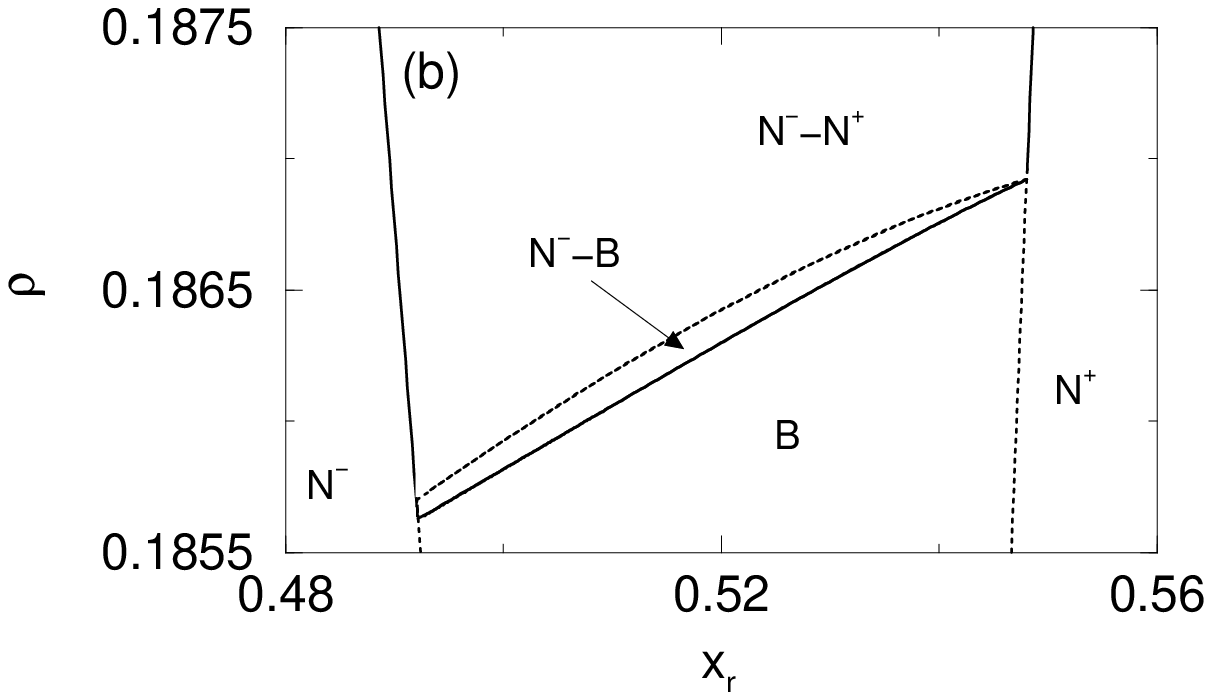}}
\caption[]{Phase diagram of a polydisperse mixture of rods and plates with 
$\kappa=5$ and length and breath polydispersities
$\Delta_L$=0.610 and $\Delta_D=0.302$. Part (b) is a zoom of the
region right above the biaxial phase.}
\label{fig5}
\end{figure}

\begin{table}
\begin{tabular}{cccc}
\hline\hline
$\alpha$ & $\Delta_L$ & $\Delta_D$ & $\kappa^*$ \\
\hline
$\infty$ & 0.000 & 0.000 & 6.9 \\
2.00 & 0.216 & 0.107 & 6.6 \\
1.00 & 0.288 & 0.143 & 6.3 \\
0.50 & 0.374 & 0.185 & 5.9 \\
0.25 & 0.470 & 0.233 & 5.1 \\
0.10 & 0.610 & 0.302 & 4.0 \\
\hline\hline
\end{tabular}
\caption[]{Threshold values of the length-to-breadth ratio, $\kappa^*$,
for the appearance of a biaxial nematic phase as a function of 
polydispersity, expressed in terms of the parameters $\alpha$,
$\Delta_L$ and $\Delta_D$.}
\label{table1}
\end{table}


We have estimated the threshold value of $\kappa$ (denoted $\kappa^*$)
beyond which the biaxial phase begins to be thermodynamically stable,
as a function of polydispersity. The results are summarized in
Table~\ref{table1}. Without polydispersity this value is $\kappa^*=6.9$,
smaller than the second virial estimation \cite{Van-Roij1}
$\kappa^*=8.8$. This is an indication that three (and higher) body
correlations increase the stability of the biaxial phase.
As we can see in Table~\ref{table1}, $\kappa^*$ decreases upon
increasing polydispersity, an illustration of the enhancement of 
stability of the biaxial ordering induced by polydispersity.

In Fig.~\ref{fig5} we plot the phase diagram of this system for
$\kappa=5$ and $\alpha=0.1$ (corresponding to $\Delta_L=0.610$ and
$\Delta_D=0.302$). According to Table \ref{table1} for this
parameter choice the biaxial phase is thermodynamically stable, and
this is clearly shown in Fig.~\ref{fig5}(a), where we can see a
small window of biaxial nematic just above the multicritical point.
Figure~\ref{fig5}(b) is a zoom of the upper border of this window. 
There is a narrow region of N$^-$--B coexistence limited from above
by a N$^-$--N$^+$ coexistence. The dividing line (the dotted line
in the diagram) represents a continuous transition between the B and
the N$^+$ phases which coexist with the N$^-$. This line has no
analog in the binary rod-plate mixture. To obtain it one has to
solve the equations for two phase coexistence at all values of
$\gamma$ in (\ref{distri}) (the volume occupied by one of the
coexisting phases) and then find the one for which the B--N$^+$
transition occurs.

The high density part of the phase diagram of Fig.~\ref{fig5}(a)
was not determined in Ref.~\cite{Martinez-Raton1}. The most remarkable
feature it shows is that pure phases are hardly stable: the diagram
is dominated by coexistence regions. The reason is that
polydispersity is so high that the mixture has, beside rods and plates,
a significant amount of cubic-like particles. This favors
entropic phase separation in the three different phases I, N$^-$ and 
N$^+$.

In the upper part of Fig.~\ref{fig5}(a) there is a region of triple
coexistence I--N$^-$--N$^+$. The lines limiting this region were
obtained by solving the coexistence equations for the case when one
of the coexisting phases is a shadow phase (i.e.\ it occupies a
vanishing fraction of the total volume). As an illustrative example,
if we take $\gamma_{\rm I}=0$, $\gamma_{\rm N^-}=\gamma$ and 
$\gamma_{\rm N^-}=1-\gamma$ and solve the coexistence equations for 
the $N=15$ unknowns (see the discussion about phase equilibria 
in polydisperse systems in Section~\ref{phase}) we obtain the curve
joining the two points marked with full squares in Fig.~\ref{fig5}(a).
The other two curves limiting the triple coexistence region
were calculated selecting $\gamma_{\rm N^+}=0$ (the left one) and
$\gamma_{\rm N^-}=0$ (the right one).

We have checked that the nematics below the triple region are always stable 
with respect to spatial density modulations of any kind. In order to
estimate the packing fraction at which these instabilities occur we
solved Eqs.~(\ref{det}) and (\ref{nabla}) using the density distributions
$\rho^{\nu}(\lambda)$ of all the coexisting phases along the borders of
the triple coexistence region. The left and right open circles correspond
respectively to the packing fractions at which the coexisting N$^+$ and N$^-$
(the shadow phase in both cases) lose their stability to the C$^+$ and
C$^-$ phases. Above these points the phase diagram will possibly include
coexistences between four phases: I, N$^-$, N$^+$ and C$^-$ or C$^+$)
(the experiments of Ref.~\cite{Kooij} find a similar scenario).

\begin{figure}
\mbox{\includegraphics*[width=3.in, angle=0]{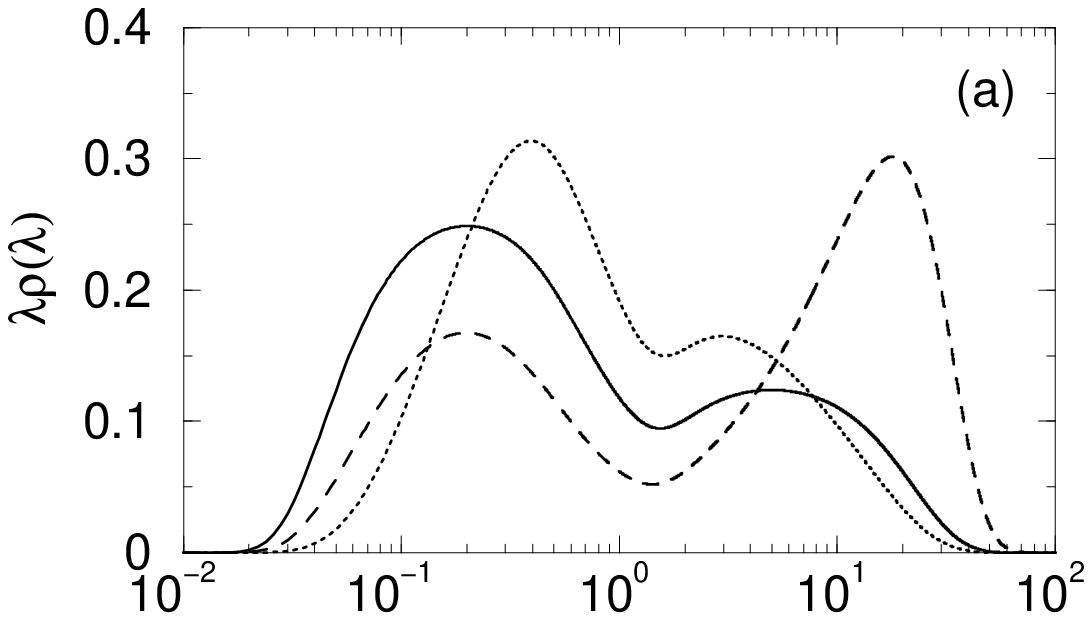}}
\mbox{\includegraphics*[width=3.in, angle=0]{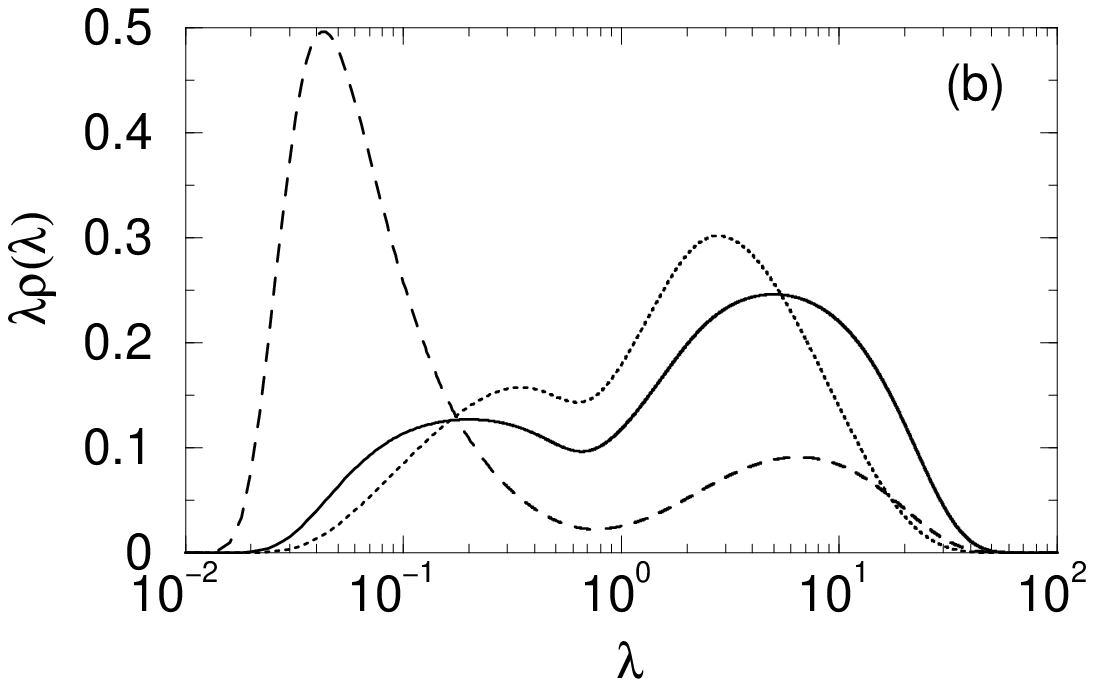}}
\caption[]{Distribution functions at the N$_{\rm sh}^-$--I--N$^+$ (a) 
and the N$_{\rm sh}^+$--I-N$^-$ (b) triple points of the phase diagram 
of Fig.~\ref{fig5} (marked with full squares). The subscript `sh'
denotes shadow phases. Solid lines represent 
the shadow phases, dotted lines the I phase and dashed lines 
the N$^+$ (a) and N$^-$ (b) phases.}
\label{fig6}
\end{figure}

Figure~\ref{fig6} shows the density distribution functions
of the phases coexisting at the points marked with
full squares in Fig.~\ref{fig5}(a). As already discussed, the curves
illustrate the high fraction of cubic-like particles partly responsible 
for the strong demixing this system exhibits. The figure also illustrates
the strong fractionation which takes place between the
coexisting phases: the isotropic phase is rich in cubic-like 
particles, the N$^-$ in plates and the N$^+$ in rods.

\begin{figure}
\hspace*{3mm}\mbox{\includegraphics*[width=3.in, angle=0]{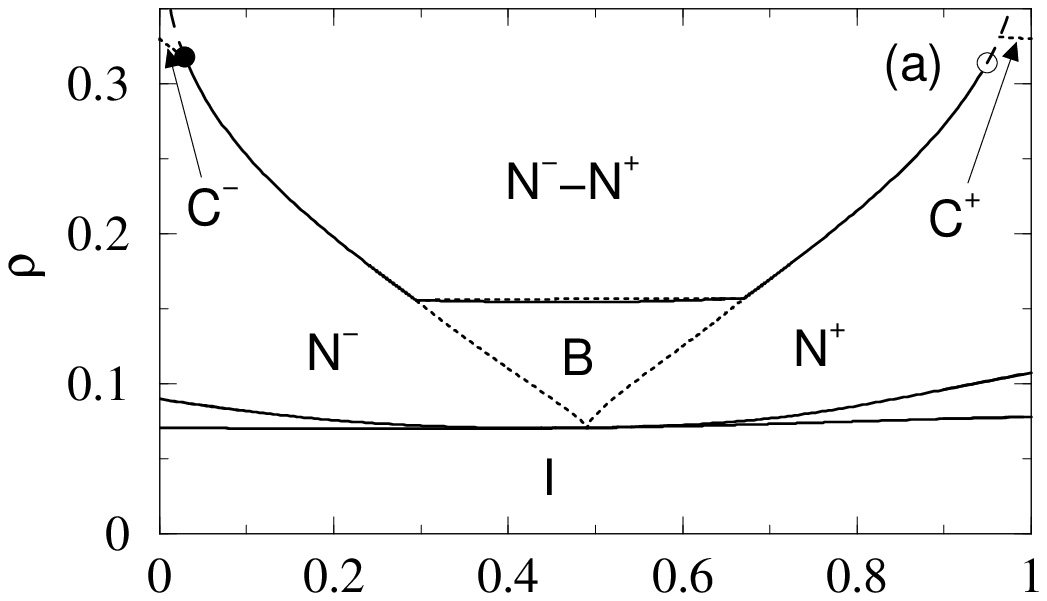}}
\mbox{\includegraphics*[width=3.3in, angle=0]{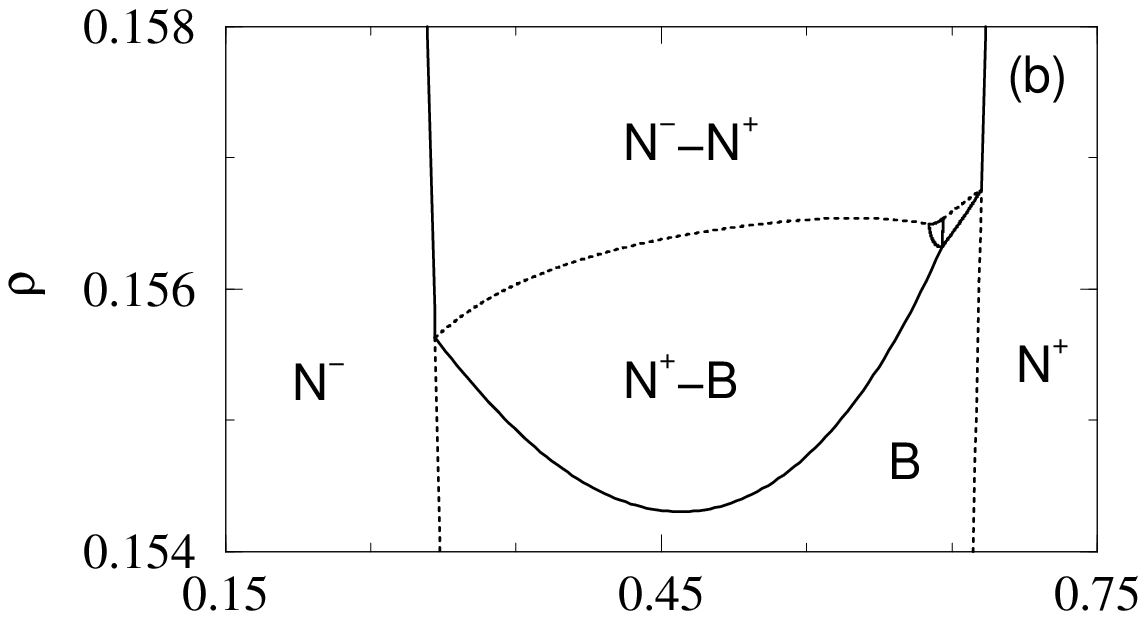}}
\hspace*{-1mm}\mbox{\includegraphics*[width=3.4in, angle=0]{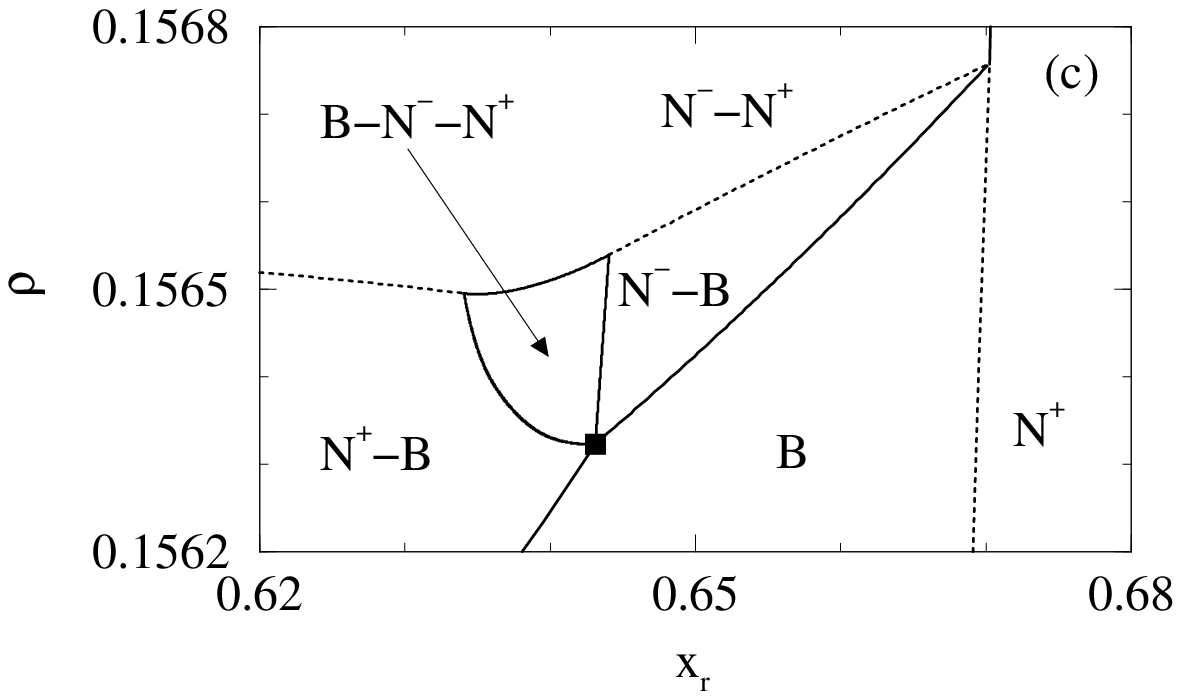}}
\caption[]{Phase diagram and two details of a polydisperse rod-plate
mixture with $\kappa=15$ and length and breadth polydispersity 
$\Delta_L=0.288$ and $\Delta_D=0.143$, respectively.}
\label{fig7}
\end{figure}

\begin{figure}
\mbox{\includegraphics*[width=3.in, angle=0]{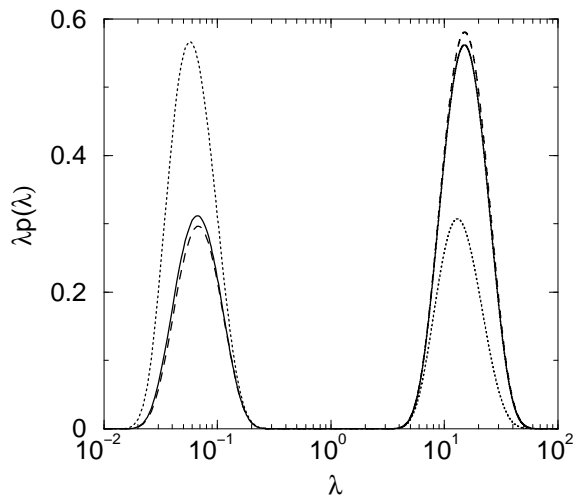}}
\caption[]{Density distribution functions at the point marked with
a full square in Fig.~\ref{fig7}(c). The solid, dashed and dotted
lines correspond to the cloud B, shadow N$^+$ and shadow N$^-$ phases,
respectively.}
\label{fig8}
\end{figure}

In contrast, the phase diagram of the polydisperse rod-plate mixture
for $\kappa=15$ does not change qualitatively compared to the binary
mixture (cf.\ Figs.~\ref{fig7} and \ref{fig4}): the size and shape of
the biaxial regions are similar and the general topology of the phase
diagram is basically the same. The only important difference (minor in
terms of the size of the portion of phase diagram it involves) appears
just above the biaxial: one can observe both B--N$^+$ and B--N$^-$
coexistence (the former occupying a larger region), as well as a
triple coexistence zone separating them [see Fig.~\ref{fig7}(c)].
As for $\kappa=5$, the B--N$^{\pm}$ coexistences become N$^+$--N$^-$
coexistence through a second order phase transition of the B phase
to the N$^{\pm}$ [dotted lines in Fig.~\ref{fig7}(b) and (c)].

The N$^-$ and N$^+$ lose their stability to columnar phases
at relative high packing fractions [see the dotted lines in the upper
part of Fig.~\ref{fig7}(a)]. In both cases it is the N$^-$ phase which
first becomes unstable with respect to C$^-$ at the
N$^+$--N$^-$ coexistence (the N$^-$ cloud on the left, marked with
a full circle in Fig.~\ref{fig7}(a), and the N$^-$ shadow on the right,
marked with an empty circle in this figure).

Figure \ref{fig8} shows the density distribution functions
corresponding to the point marked by a full square in Fig.~\ref{fig7}(c).
The distributions of the N$^+$ and B phases are very similar, except
for the fact that the N$^+$ has a slightly higher proportion 
of rods (and correspondingly less of plates) than the $B$ phase.
It is clear from the figure that the proportion of cubic-like 
particles is negligible, so this mixture can be regarded as a true
(polydisperse) rod-plate mixture.

\section{Conclusions}
\label{conclusions}
We have studied the effect of polydispersity on the phase diagrams 
of several variants of the Zwanzig model for liquid crystals. We have
first determined the spinodal instabilities of the homogeneous phases to 
the smectic, columnar or plastic solid phases for one component fluids 
of rods or plates. There is a general qualitative agreement with 
phase diagrams of oblate and prolate particles obtained from 
Monte Carlo simulations \cite{Bolhuis1,Veerman}. We have also shown
that even a small degree of polydispersity has dramatic effects on
the transitions to highly ordered phases (like the smectic or the
columnar). In particular, we have shown that upon increasing
polydispersity the transition to the columnar phase of rods 
preempts the N$^+$--S$^+$ transition, in agreement with simulations
\cite{Polson}.

Next we have obtained the phase diagrams of symmetric rod-plate binary
mixtures with length-to-breadth ratios $\kappa=5$ and 15. While for
large $\kappa$'s the differences between these phase diagrams and
those calculated with a second virial approximation \cite{Van-Roij1}
are only quantitative, for small $\kappa$'s both methods differ
qualitatively. For instance, all the transitions occur at packing 
fractions well below those obtained with the second virial approach,
and sometimes the relative stability between different inhomogeneous
phases changes.

We have shown that the introduction of polydispersity in a rod-plate 
mixture in a very symmetric way enhances the stability of the biaxial 
phase even for as small an aspect ratio as $\kappa=5$. For this case 
the amount of polydispersity needed to stabilize the biaxial phase is
so high that the mixture becomes very unstable with respect to phase
separation at practically any composition. There are large regions of three 
phase and possibly four phase coexistence (I--N$^-$--N$^+$ and 
I--N$^-$--N$^+$--C$^{\pm}$). These results agree with what is observed
in experiments with rod-plate mixtures with a high degree of
polydispersity \cite{Kooij}.

\begin{acknowledgments}
This work is part of the research Project No. BFM2000-0004 (DIGI) of the 
Ministerio de Ciencia y Tecnolog\'{\i}a (Spain). Y. M-R. is supported 
by a postdoctoral grant of the Consejer\'{\i}a de Educaci\'on de la 
Comunidad de Madrid.
\end{acknowledgments}

\end{document}